\documentclass[aps,prl,twocolumn,showpacs,groupedaddress]{revtex4-1}
\usepackage{amsmath}

\usepackage{graphicx}
\usepackage{dcolumn}
\usepackage{bm}
\usepackage{amssymb}
\usepackage{dcolumn}
\usepackage{bm}
\usepackage{verbatim}

\begin{document}

\title{Patterns, Shapes and Growth of a Spherical Superabsorvent Polymers}
\author{M. Kyotoku$^1$}
\author{E. Nogueira Jr.$^1$}
\author{B.B.C. Kyotoku$^2$}
\author{C. Salvador$^1$}
\affiliation{$^1$Departamento de F\'{\i}sica, Universidade Federal da Para\'{\i}ba }
\affiliation{$^2$ Departamento de Eletr\^{o}nica Qu\^{a}ntica,Instituto de F\`{\i}sica Gleb Wataghin, 
 Universidade Estadual de Campinas
             }
\date{\today }

\begin{abstract}
\noindent Combining experimental procedures and theoretical modeling, we
studied patterns, shapes, and growth of a special type of
SuperAbsorbent Polymers (SAP). At the initial stage, embed in water, we have a polygonal pattern over a
spherical form, that evolves into a wrinkled deformed shape, which is then
followed by an oblate ellipsoid with some of wrinkles over the surface. The final step
is a slowly growing sphere. We measured the weights in function of time during the growing process. 
The results can be represented by an exponential function, which is derived from the general continuity equation.
\end{abstract}

\pacs{61.41.+e, 05.70.Ln, 68.10.-m, 82.70.Gg}
\maketitle
In the present letter, we report an integrated research in patterns,
shapes and growth, as suggested by D'Arcy Thompson in
his celebrated book \textquotedblleft On the Growth and
Form \textquotedblright \cite{Thompson} . This growth is mathematically analyzed with the
continuity equation as we embed in water, a very complex network
of cross linked hydrophilic polymers called SuperAbsorbent Polymer(SAP).

In general, the SAPs are commonly made from the polymerization of acrylic acid 
blended with sodium hydroxide.  As we know the SAP is kind of gel studied \cite{Hajime Tanaka},\cite{Suematsu} in similar scenery decades ago.  Due to the their high absorbency, the SAPs have been used in diapers 
as well as water reservoir for agriculture \cite{iranianos} and
horticulture and the latest accomplishment in this field is the confection
of a superabsorbent membrane called Imec \cite{Imec}.
There is also proposals to be applied in high-performance concrete
\cite{concret} and petroleum \cite{petroleum}.

In the early years of the present decade. SAP was retooled and sold
world wide as a tiny hard spheres, with a diameter of around 3.5 mm, while
the original one has an irregular shape, similar to grains or even a white powder.

\begin{figure}[h]
\centering
\includegraphics[scale=0.60]{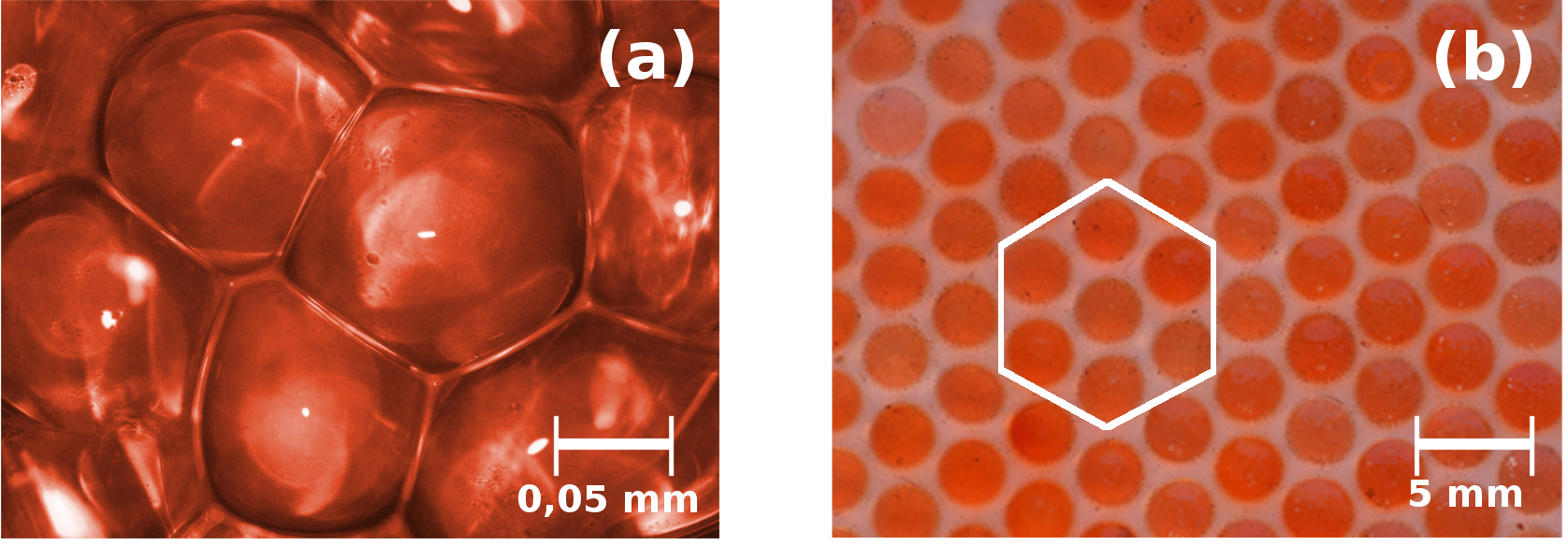}  
\caption{(a) Surface polygonal pattern emerging after seconds in water. (b)
Hexagonal patterns in water, produced by a random distribution of SAP spheres. }
\label{hexagonal.near}
\end{figure}

One of these SAP spheres or Crystal Gel is immersed in water for a
few seconds. Removing and looking at it through a microscopy, we see a 
polygonal pattern as shown in Fig.\ref{hexagonal.near}(a). The origin
of this pattern can be traced back to the interaction between the water
dipole with the hydrophilic polymer. This guess can be verified using the
following procedure.

\newpage

SAP spheres were left randomly in a small container. Then, few second in the
water, as noted in Fig. \ref{hexagonal.near}(b) the SAP spheres become a
hexagonal pattern, and the equidistance between them shows the propagation
of the dipole interaction through the water. The polygonal irregularity is
mainly due to the shielding of the dipole-dipole interaction by the medium.
We need to recall that a Buckyball effect has some role in this
non-uniformity \cite{Buckyball}.

\begin{figure}[h]
\includegraphics[scale=.34]{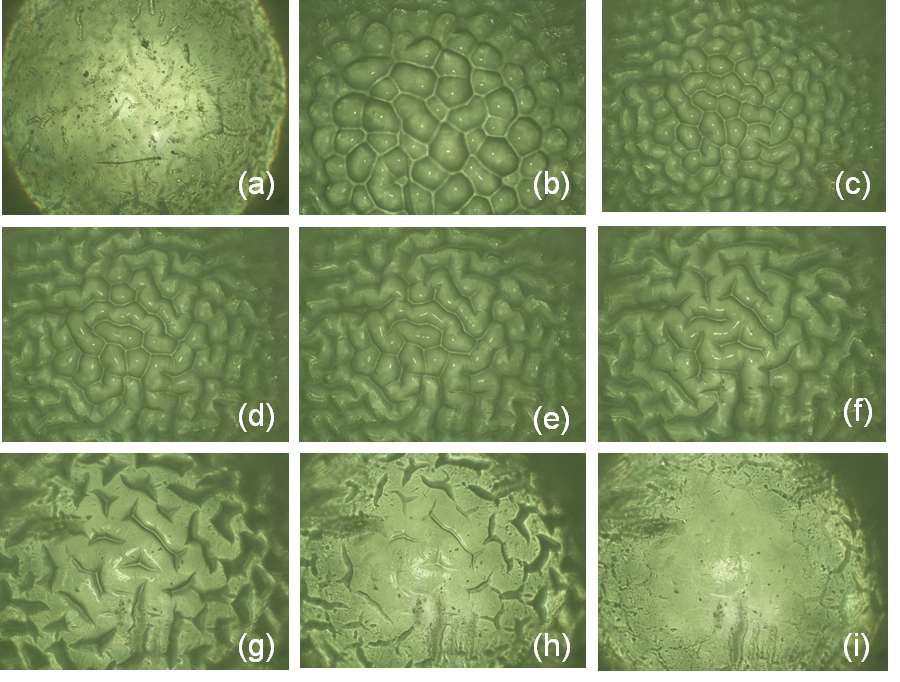}  
\caption{Sequence shows the smoothing process of the spherical SAP after
submersion and leaded outside water. (a) Shows a dry Crystal Gel before the dive.
(b) to (i) time evolution about two minutes. Some of these frames
looks like a human brain surface. }
\label{mosaic}\centering
\end{figure}

Aside of polygonal pattern another fact can be seen under the microscope.
Out of water, the polygons coalesces, as can be observed in a photo sequence
in Fig. \ref{mosaic}. In the middle part, we notice a labyrinthine pattern
similar to the surface of our brains. This phenomenon seems to be analogous
to the work on surface pattern of fruit dehydration \cite{surface}, where
they examined the elastic interaction between the skin and the fruit pulp.
The smoothing of surface, noticed in the Fig.\ref{mosaic}, is due to the
homogenization through the water diffusion over the entire surface.

\newpage

\begin{figure}[h]
\centering
\includegraphics[scale=0.65]{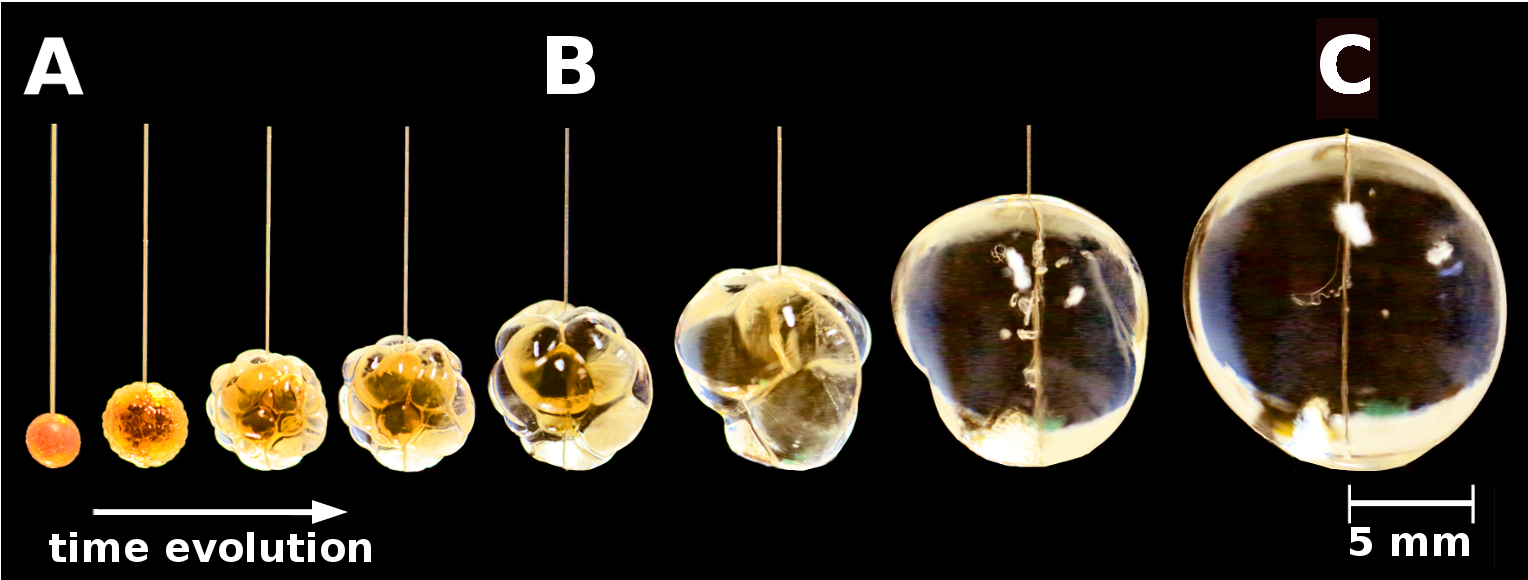}  
\caption{ Patterns, shapes and growth of the same face of sphere outside
water. (A) Dry sphere. (B) a shape transition around twenty minutes can be
noticed. Finally, (C) spherical shape, after twenty four hours. The fine
needle supports the SAP sphere.}
\label{Growth}
\end{figure}

In order to look for different forms and patterns, the SAP sphere was submerged, sequentially, in a
distilled water with acidity around pH 4.0. For this purpose, we built an
automated system, supporting the ball with very fine needle and, taking it
out time to time, in order to take a picture. This step was
needed, since the water refraction does not allow a good picture. The dives
were three minutes inside water and ten seconds out of it. This procedure
ensures the surface to be wet during the photography stage. Fig.\ref{Growth} 
shows different shapes of SAP sphere were only the
relevant pictures is displayed.

Polygonal patterns over spherical form can be seen in Fig.\ref{Growth} 
for around the first twenty minutes with water diffusing through
the skin. After that, as the water enters little deeper into the polymer and 
the shape transformation occurs. The Crystal Gel becomes highly deformed and 
the surface pattern changes, it smooths and a scar or an elastic folding 
appears over the surface.

If we take out the highly deformed Crystal Gel from the water, the surface
self smooth and after some time will take a spherical shape. In this
point another experiment was performed, but not displayed. Using a razor we
cut the deformed Crystal Gel in two parts, both twists considerably 
and after some time they can be re-assembled into a single sphere. 

Before reaching the ultimate spherical form, an oblate ellipsoidal shape 
with a cross and semi-cross elastic folding can be noticed. Eight hours
after, the Crystal Gel is spherical and the growing
slows and finally a saturated phase is reached after around thirty hours.

All these experiments shows, the
diffusion  character of these phenomena, the search for homogenization. 
In the final stages, the concentration of water 
inside and outside the sphere is practically the
same, with a dynamic equilibrium state. 

Since, the Crystal Gel spheres are commercialized in different colors, we 
experimentally noted, that the overall behavior does not depend on the 
pigment material used to color these spheres.

The next step is to measure the growing process.
We perform an experiment by selecting carefully, five sets with ten tiny
balls and weighting them with a balance with a scale of hundredth grams of
precision and, we show them in the first row of  Tab.\ref{tabela1}. 
With the help of a sieve, each set was dived multiple times into a water with 
same acidity earlier mentioned and the weight was measured . 

\begin{table}[h]
\begin{ruledtabular}
\begin{tabular}{cccccc}
Set & 1 & 2 & 3 & 4 & 5\\ \hline
Multiple dives & 0,46 & 0,48 & 0,46 & 0,47 & 0.46 \\
Single dive & 0,45 & 0,46 & 0.44 & 0,43 & 0,45\\ 
\end{tabular}
\caption{Weight in grams of each set of 10 spheres. 5 sets were used in multiple dives experiments and 5 sets were used in single dive experiments.}\label{tabela1}
\end{ruledtabular}
\end{table}

\begin{figure}[h]
\centering
\includegraphics[scale=0.46]{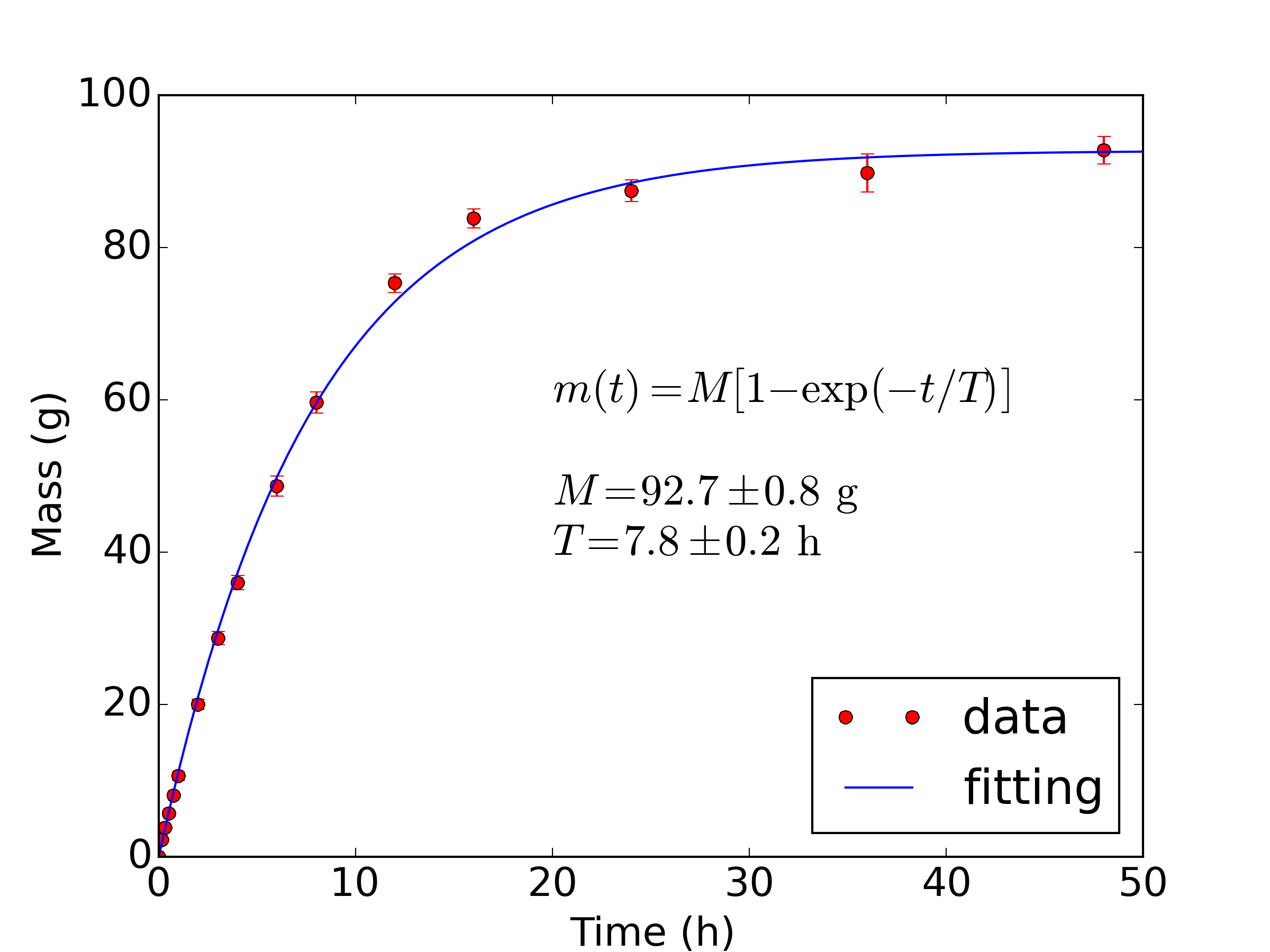}  
\caption{ 
Mass of absorbing water in function of time to five sets where
each has ten Crystal Gels. The red dots are the weight's experimental
results.}
\label{grafico}
\end{figure}

\begin{figure}[h]
\includegraphics[scale=.5]{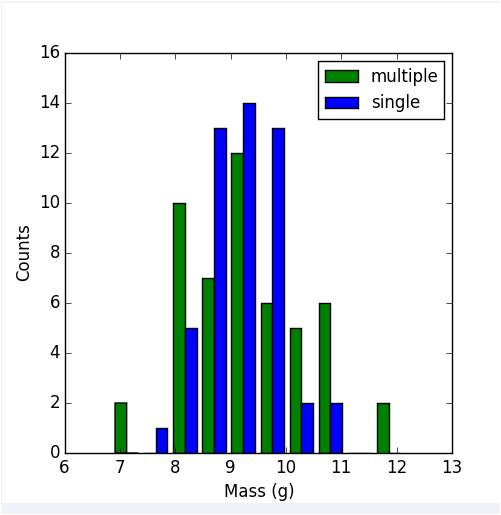}   
\caption{ The comparison between the counts of individual weight spheres, in a single dive experiment - blue color - and
multiple dive experiments - green color. Note the different dispersion}
\label{dispersion}
\centering
\end{figure}

Results are displayed as a dots in the Fig.\ref{grafico}. Time variation
was measured with Nikon DX5100 camera clock and simultaneously the weight
measurement was taken. Ending this experiment, we weight the individual spheres and
the results will be discussed in Fig.\ref{dispersion}. 

As you can notice, an exponential curve fits, with the experimental data,
the error bar cannot be seen in the beginning of the curve, however it
becomes larger and larger. Most likely the multiple dives, scratches the
surface, allowing more water in the final steps. 

To test this hypothesis, a second measurement of the final weight was done 
using a different batch of 5 sets of spheres, see the second row of Tab.\ref{tabela1}. 
But this time the spheres were dived only once and no measurement was made as they 
were growing. In
 Fig.\ref{dispersion} shows the mass distribution of all the 
final spheres weight. Blue column represents one dive and green color, multiple dives. 
As expected in one dive, the dispersion is lower compared to multiple dives.

In the present experiment the flux of water, is going after, as we already mentioned, for homogenization. Therefore we have, only,  a local form of conservation law which is governed by the general second Fick's Law equation. 
Here written in the integral form:

\begin{equation}
\int \frac{\partial \rho _{\text{H}_{\text{2}}\text{O}}\left( 
\overrightarrow{r},t\right) }{\partial t}dV+\oint \overrightarrow{J}\bullet 
\overrightarrow{n}dS=0, 
\label{Fick}
\end{equation}

\noindent where $\rho _{H_{2}O}\left( \overrightarrow{r},t\right) $
represents the density or the concentration of water inside the Crystal Gel,
which increases due to absorption. The $\vec J $ is the flux of water going
towards the center of the sphere and is defined by the following expression:

\begin{equation}
\overrightarrow{J}=D\left( \rho _{\text{H}_{\text{2}}\text{O}},%
\overrightarrow{r}\right) \overrightarrow{\nabla }\rho _{\text{H}_{\text{2}}%
\text{O}}\left( \overrightarrow{r},t\right). 
\label{J}
\end{equation}

The function $D\left( \rho _{\text{H}_{\text{2}}\text{O}},\overrightarrow{r}
\right) $ is the Diffusion Coefficient. Physically, for each radius it may
describe the permeability of the surface. An assemble of these surfaces is 
a collection of tortuous micro pipes with different diameters. As the Crystal 
Gel swallows the water, these micro pipes enlarges and we have in each of them, a
different concentration of water.

The streaming of water inside the Crystal Gel is described 
by the gradient of concentration $\overrightarrow{\nabla }
\left( \rho _{_{\text{H}_{\text{2}}\text{O}}},\overrightarrow{r}\right) $.
As mentioned, experimentally in the first twenty to twenty five minutes, it
seems to be a diffusive spherical field. After this time, as the streams of
water overcame the skin or the first layer and the water dipoles penetrates
deep in the second complex cross linked polymer network, and two types of
elasticity seems to appear, one in the skin another in the pulp and the
interaction between them may explain the overall aspect of the large
deformations displayed {Fig.}\ref{Growth}.

At this moment, it is not easy to guess a reliable density dependent
diffusion coefficient. They should incorporate the permeability and the
elasticity of the gel and, simultaneously, the shape deformation evolution.
Further, it should also incorporate the leveling of gel surface.

Additionally, if one remembers the differential form of Eq.\ref{Fick},
this coefficient is coupled through the divergent operator, with the
gradient of water concentration, leading to non linear Differential
Equation, and by solving it, someone will describe the concentration of water 
evolution,  as it is displayed in the {Fig. }\ref{Growth}.

However, we do not have enough material to pursue this ambitious and
interesting research program. Fortunately, as a first step, we can interpret
the second term of the Eq.\ref{J} as the flux of water $%
\overrightarrow{J}$ into the surface. Then, let us assume, that is always possible
to transform the complicated shapes observed in the Fig.\ref{Growth} into
spheres and the water diffuses radially. In others words we have field
similar to a central field and is:

\begin{equation}
\overrightarrow{\nabla }\rho _{\text{H}_{\text{2}}\text{O}}\left( 
\overrightarrow{r},t\right) =\widehat{r}\frac{\partial \rho _{\text{H}_{%
\text{2}}\text{O}}\left( r,t\right) }{\partial r}. 
\label{equ3}
\end{equation}

\noindent Replacing this term into the second term of the Eq. \ref{Fick} and 
performing the surface integral we get,

\begin{equation}
\begin{split}
\int \frac{\partial \rho _{\text{H}_{\text{2}}\text{O}}\left( r,t\right) }{%
\partial t} dV-D(\rho _{\text{H}_{\text{2}}\text{O}},r_{\text{surf}})& \\
\frac{ \partial \rho _{\text{H}_{\text{2}}\text{O}} \left( r_{\text{surf}%
},t\right) }{\partial r_{\text{surf}}}4\pi r_{\text{surf}}^{2}=0.
\label{equ4}
\end{split}%
\end{equation}

The radial component of the gradient in the above Eq.\ref{equ4} can be interpreted as the flux of
water penetrating into the surface with radius $r_{\text{surf }}$. Consequently the density of
water decreases from outside to inside of the sphere in a radial and uniform
way. Mathematically this is expressed by,

\begin{equation}
\frac{\partial \rho _{\text{H}_{\text{2}}\text{O}}\left( r_{\text{surf}},t\right) }{
\partial r_{\text{surf}}}\cong \frac{\rho _{%
\text{H}_{\text{2}}\text{O}}-\rho _{\text{H}_{\text{2}}\text{O}}\left( r_{%
\text{surf}},t\right) }{\Delta r_{\text{surf}}}.
\label{equ5}
\end{equation}

Let us now discuss the Diffusion Coefficient from Eq.\ref{equ4}. 
From our experimental data, we can notice that the water
flux is at maximum when polymer is placed in water and as the time evolves, the
absorbency capacity diminishes and finally reaches the some kind of steady stage. 
The amount of water going in and out is
practically the same, a dynamical equilibrium.

This dynamical situation can be abridged considering a net permeability. In
our case, it can be represented by an association with both, the thickness $\Delta r_{\text{surf}}$
and the inverse of spherical surface i.e., $\Delta r_{\text{surf}}/4\pi r_{\text{surf}}^{2}$.
Furthermore, an additional constant coefficient $\varepsilon \ $is needed to classify,
simultaneously, the characteristic of Crystal Gel and the embedding medium. 

Thus we suggest, rename $D\left( \rho _{\text{H}_{\text{2}}\text{O}%
},r_{surf}\right) $ as effective diffusion coefficient and we make a
following educated guess,

\begin{equation}
D_{\text{eff}}=\frac{\Delta r_{\text{surf}}}{4\pi \varepsilon r_{\text{surf}%
}^{2}}.
\label{equ6}
\end{equation}

Inserting, both, the effective diffusion coefficient and Eq.\ref{equ5} into the Eq.\ref{equ4} we get:
 
\begin{equation}
\varepsilon \int \frac{\partial \rho _{_{\text{H}_{\text{2}}\text{O}}}\left(
r,t\right) }{\partial t}dV-\rho _{\text{H}_{\text{2}}\text{O}} +\rho
_{H_{2}O}\left( r_{\text{surf}},t\right) =0.
\label{equ7}
\end{equation}

Now we need to perform the calculation about the volume integral in this equation.
As we do not know the variation of the of water inside the sphere,
let us assume, that for each time interval, the water drips on a fixed volume, 
instead of adopting a Crystal Gel expansion, a real
physical situation. 

So the evolution of the density of water will be
independent of the radius of the sphere, but there will be variation of
concentration. We will also adopt the fixed volume as the maximum volume
reached when saturation occurs or $V_{\text{max}}$. Thus variation of water
concentration is,

\begin{equation}
\frac{\partial \rho _{\text{H}_{\text{2}}\text{O}}\left( r,t\right) }{%
\partial t}=\frac{1}{V_{\max }}\frac{dm_{\text{H}_{\text{2}}\text{O}}(t)}{dt}%
.
\end{equation}

Replacing it, on the first term of the continuity equation Eq.\ref{equ7} and performing
the integration over the volume, we obtain the well known differential
equation,

\begin{equation}
\rho _{\text{H}_{\text{2}}\text{O}}-\varepsilon \frac{dm_{\text{H}_{\text{2}}%
\text{O}}(t)}{dt}-\frac{m_{\text{H}_{\text{2}}\text{O}}(t)}{V_{\text{max}}}%
=0,
\end{equation}
which results in:

\begin{equation}
m(t)=V_{\text{max}}\rho _{\text{H}_{\text{2}}\text{O}}\left[ 1-e^{-\frac{t}{%
\varepsilon V_{\text{max}}}}\right].  \label{equ12}
\end{equation}

This function describes perfectly our measurement displayed in
Fig.\ref{grafico},  reassuring that all the spherical approximations made in the integral form of the second Fick's Law, is robust and reliable.  We can further blend it with fitted data displayed in Fig.\ref{grafico} to get $ V_{\text{max}} = (92.7\pm 0,8) ml $ and $\varepsilon= (8,4 \pm 0,2) 10^{-2}$ h/ml, where we assumed $\rho _{H_{2}O}\cong 1.0 $ g/ml.

However, as we mentioned, the real picture is much more complicated.  Additionally,
the inclusion of the angular terms of the spherical coordinated is needed and also, 
a more realistic Diffusion Coefficient.

These additions will furnish a non-linear Differential Equation and, its solution will reveal 
a complete interpretation of the evolution of patterns, shapes and growth displayed throughout the pictures
of present paper. Furthermore, this approach will also reveals the water evolution inside the Crystal Gel which
can be subjected to an experimental scrutiny by cutting the Crystal Gels in
different stages of its evolution.

Finally, as we mentioned, we developed a methodology  to investigate
in an integrated manner, patterns, shapes and growth of SAP spheres. 
Since, our approach results in the core of the Ontogenetic Growth
Model, \cite{ontogenetic}, which fits the grow of several animals, 
where the Energy is the diffusing material. Then, it is tempting
to suggest that present procedure could be used to derive their basic 
differential equation from the general Continuity Equation.


\begin{thebibliography}
\bibitem{} 
\bibitem{Thompson} D. W. Thompson, On Growth and Form,(Cambridge University Press,  1992). 
\bibitem{Hajime Tanaka} H. Tanaka, H. Tomita, A. Takasu, T. Hayashi, and T. Nishi,  Phys.Rev.Lett., \textbf{68}, 2794 (1992).
\bibitem{Suematsu} N. Suematsu, K. Sekimoto and K. Kawasaki, Phys.Rev A - Rapid Communications, \textbf{41}, 5751 (1990).

\bibitem{iranianos} J. M. Zohuriaan-Mehr and K. Kabiri, Iranian
Polymer Journal, \textbf{17}, 6, 451 (2008).

\bibitem{Imec} Y. Mori, Reactive and Functional Polymers, \textbf{73}, 7,
936 (2013).

\bibitem{concret} O. M. Jensen, Concrete International, \textbf{35},
1, 48 (2013).

\bibitem{petroleum} J. Alaei, S.H. Boroojerdi and Z. Rabiel, Petroleum and Coal \textbf{47}, 32 (2005).

\bibitem{Buckyball} H. W. Kroto, J. R. Heath, S. C. O'Brian, R. F. Curl and R. E. Smalley Nature, \textbf{318}, 6042 (1985).

\bibitem{surface} B. Li, F. Jia, Y.P. Cao, X. Q. Feng, H. Gao  Phys.Rev.Lett., \textbf{106}, 234301 (2011).

\bibitem{ontogenetic} G. B. West, J. H. Brown and B. J. Enquist, Nature, \textbf{413}, 628 (2001).

\end{thebibliography}
\end{document}